\documentclass[twocolumn,showpacs,preprintnumbers,amsmath,amssymb]{revtex4}
\usepackage{graphicx}% Include figure files
\usepackage{dcolumn}% Align table columns on decimal point
\usepackage{bm}% bold math

\begin{document}

\preprint{Appl. Phys. Lett. {\bf 88}, June (2006), in press.}

\title{Drastic improvement of surface structure and current-carrying ability in YBa$_2$Cu$_3$O$_7$ films by introducing multilayered structure}

\author{Alexey V. Pan}
\email{pan@uow.edu.au}
\author{Serhiy Pysarenko}
\author{Shi X. Dou}
\affiliation{Institute for Superconducting and Electronic Materials, University of Wollongong, \\ Northfields Avenue, Wollongong, NSW 2522, Australia}

\date{November 19, 2005}

\begin{abstract}
Much smoother surfaces and significantly improved superconducting properties of relatively thick YBa$_2$Cu$_3$O$_7$ (YBCO) films have been achieved by
introducing a multilayered structure with alternating main YBCO
and additional NdBCO layers. The surface of thick (1~$\mu$m)
multilayers has almost no holes compared to
YBCO films. Critical current density ($J_c$) have been
drastically increased up to a factor $>3$ in 1~$\mu$m multilayered structures
compared to YBCO films over entire temperature and applied
magnetic filed range. Moreover, $J_c$ values measured in thick
multilayers are even larger than in much thinner YBCO films. The
$J_c$ and surface improvement have been analysed and attributed to growth
conditions and corresponding structural peculiarities.
\end{abstract}

\pacs{74.78.Bz, 74.78.Fk}

\maketitle

\section{Introduction}

Various superconducting applications demand different forms of the high temperature superconducting (HTS) films in terms of thickness, composition ({\it e.g.} multilayers), properties and performance. To establish a technology for growth of high quality YBa$_2$Cu$_3$O$_7$ (YBCO) films with a single-crystal structure, an enormous effort has been made. However, a clear solution to some important problems has not been found. For example, properties of thick YBCO films for coated conductors significantly degrade as the thickness of the films is increased over about 400~nm; considerable surface roughness of YBCO films and their growth problems over large-angle boundaries on bicrystals obstruct the development of reproducible Josephson junctions (JJs) of different types. 

In this work, we show that films with relatively large thicknesses $\sim 1$~$\mu$m can exhibit electromagnetic and structural properties, which outperform thinner films with ``optimal" thicknesses if a multilayered structure is introduced. 

It is well known that superconducting films generally exhibit the inverse dependency of the critical current density ($J_c$) on their thickness ($d_p$): $J_c \propto 1/d_p$ \cite{bee02,kho04,fol99,pan05}. This dependence is usually valid starting from a certain optimal thickness. At thicknesses smaller than the optimal one, $J_c$ drops much more rapidly than at larger thicknesses (Fig.~\ref{thick}). This $J_c$-$d_p$ relation is slightly technique dependent. For example, in the case of pulsed laser deposition (PLD) the optimal thickness may vary approximately from 50 to 400~nm, depending on a certain set of deposition parameters ({\it e.g.} deposition rate) \cite{pan05}. In any case, the $J_c(d_p)$ dependence usually has a maximum at the optimal thickness. Hence, for coated conductors, it is desirable to grow thicker films, preserving $J_c$ of the films with the optimal thickness.

It is also well established that the surface roughness and single-crystalline structure of the films degrade as the thickness of the films increases \cite{cha90}. Gain in smoothening of the surface roughness is minimal for films deposited at slower rates \cite{pan05,cha90}. Larger surface inhomogeneity reduces total supercurrent and obstructs fabrication of various types of JJs.

It has been shown \cite{ham00,cai04,jia02,jia03} that introducing various multilayered structures with alternating layers can positively influence superconducting performance. However, only relatively thin films $< 1 \, \mu$m have been investigated, or no influence on the surface morphology has been shown.

In this work, we show that if YBCO films are deposited in the form of ReBCO superconducting multilayers (where Re is a rare earth element), the film properties are significantly improved in terms of the $J_c(B_a)$ performance, as well as surface roughness and overall homogeneity of the films. In fact, we have found that YBCO/ReBCO multilayers (with Re = Nd-element) of about 1~$\mu$m thick can outperform not only YBCO mono-layer films of the similar thickness, but also YBCO films with optimal thicknesses.

\section{Experimental Details}

\begin{figure}%[t]
%\begin{center}
%Requires \usepackage{graphicx}
\vspace{-1.0cm}
\hspace{-0.5cm}
\includegraphics[width=80mm]{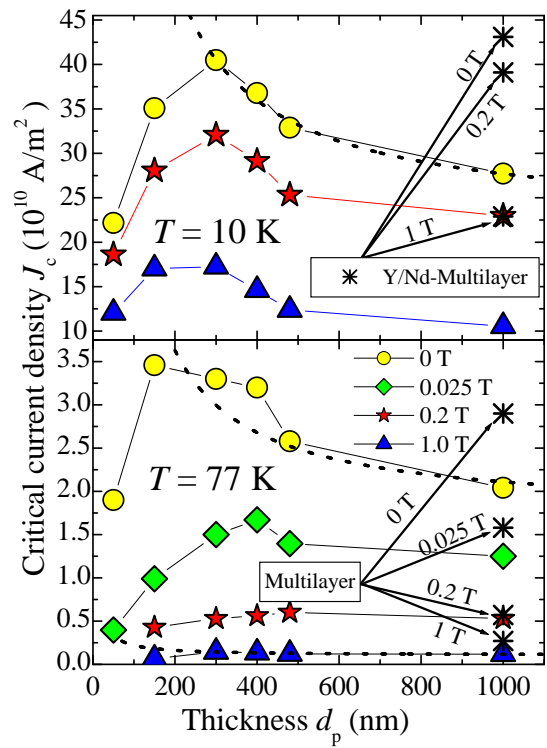}
\vspace{-1.0cm}
\caption{\label{thick}$J_c$ dependence on the film thickness ($d_p$) at 10~K (a) and at 77~K (b) in different fields. The dotted lines show $J_c \propto d_p^{-1}$ fits. The asterisks show $J_c$ values for the multilayers in the corresponding fields.}
%\end{center}
\end{figure}

High quality YBCO films and YBCO/NdBCO multilayers have been grown by pulsed-laser deposition with the help of KrF Excimer Laser (248~nm) on (100) SrTiO$_3$ (STO) substrates in oxygen atmosphere of 40~Pa. The distance between YBCO target and substrates was about 5~cm. The optimal deposition temperature (at which the highest $J_c(0, 77 \, {\rm K})$ is obtained) for the YBCO films was found to be 780$^{\circ}$C. The optimal thickness (with the same criterion as for the deposition temperature) was found to vary from 0.1 to 0.4~$\mu$m (Fig.~\ref{thick}). Our main interest was to improve characteristics, such as surface morphology and $J_c$, of thicker YBCO films of about 1~$\mu$m thick by introducing alternating layered structure. Therefore, we have prepared a series of YBCO/NdBCO multilayers with the resultant thickness equal to 1~$\mu$m thick YBCO film. The multilayered structure considered in this work is as follows. A 300 nm thick YBCO layer is deposited directly on the substrate, then a 50~nm thick NdBCO film is deposited, followed by a 300~nm YBCO layer, a 50~nm NdBCO, and a 300~nm YBCO layers. By scanning electron microscopy (SEM), it is possible to see both NdBCO layers in the image with the highest magnification shown in Fig.~\ref{sem}(d).

The optimal deposition temperature for monolayer NdBCO films was established to be about 50$^{\circ}$C higher than that for YBCO films. Therefore, we have varied the deposition temperature of the multilayer to find its optimal growth conditions. It turned out that the optimal deposition temperature for the multilayers is nearly the same as for pure YBCO films. Moreover, the temperature range for obtaining high quality multilayers is wider than for pure YBCO films \cite{pan05}. The critical temperature ($T_c$) for both YBCO films and the multilayers varies slightly from 89.3~K to 91.7~K with the transition width of $< 0.5$~K measured by AC susceptibility.

The surface morphology of the films have been observed at small angles ($\sim 10^{\circ}$ to 20$^{\circ}$) to the plane of the surface with the help of SEM. Electromagnetic properties of the films have been investigated by magnetization measurements over a wide applied field ($|B_a| \le 5$~T) and temperature (5~K $\le T \le 95$~K) ranges. DC magnetic fields have been applied perpendicular to the film plane. $J_c(B_a, T)$ dependences have been obtained from the width of the magnetization loops, using the critical state model: $J_c = 2 \Delta M /[w_p(1 - w_p/3l_p)]$ in A/m$^2$, where $\Delta M = |M^+| + |M^-|$ taken from magnetization loops measured at different temperatures versus the applied field, $w_p$ and $l_p$ are respectively the width and length of the films measured.

\section{Results and Discussions}

\begin{figure}%[t]
%\begin{center}
%Requires \usepackage{graphicx}
%\vspace{3.5cm}
%\hspace{-6.0cm}
\includegraphics[width=80mm]{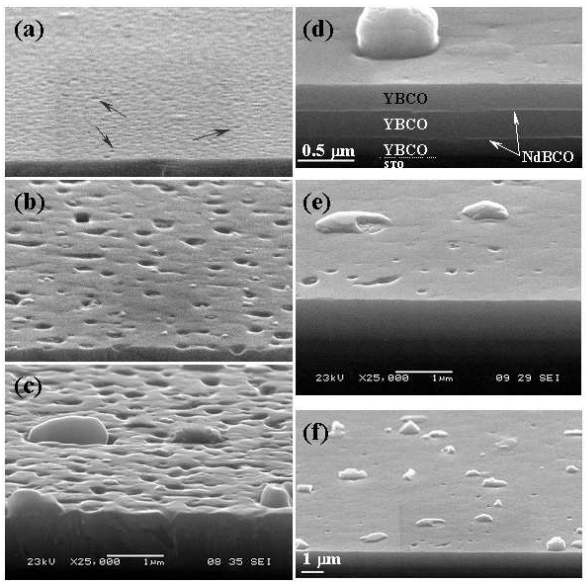}
%\vspace{-5.0cm}
\caption{SEM micrographs of the surface morphology of different YBCO films being (a) 0.1~$\mu$m, (b) 0.4~$\mu$m and (c) 1~$\mu$m thick (the scale bar valid for all images in the left column is provided at the bottom of (c)), as well as one of the YBCO/NdBCO multilayers at different magnifications (d), (e), (f), so that (e) has the same magnification as (a), (b), and (c) for comparision. At the highest magnification for the multilayer shown in (d), three thick YBCO layers separated by two thin NdBCO layers (shown by the arrows) can be observed. The dotted line in (d) indicates the STO/YBCO interface. The holes are clearly seen in (b) and (c), whereas in (a) some of the holes positions are shown by the arrows.}
\label{sem}
%\end{center}
\end{figure}

In Fig.~\ref{sem}, we show a typical surface of a 1~$\mu$m thick YBCO/NdBCO multilayer at different scales (Fig.~\ref{sem}(d),(e),(f)), as well as YBCO films having thicknesses of 0.1~$\mu$m, 0.4~$\mu$m and 1~$\mu$m (Fig.~\ref{sem}(a),(b),(c), respectively). Strikingly, the multilayer exhibits an {\em extremely smooth} surface (Fig.~\ref{sem}(e)) with only few hole-like features, very few droplets (or outgrowths), and no sign of even slight ``bumpiness", which is observed for the 0.1~$\mu$m film (Fig.~\ref{sem}(a)). In contrast to the multilayer, the YBCO film with the same thickness 1~$\mu$m (Fig.~\ref{sem}(c)) has numerous holes, which are characteristic for the spiral growth of this composition. The holes ($\sim 0.2$~$\mu$m diameter on the surface) and corresponding structural inhomogeneities extend throughout the entire thickness of the film. In addition, some droplets (outgrowths) up to 1~$\mu$m large can be found on the surface of the thickest film. The resultant surface appears to be very rough. Smoother surfaces are obtained for 0.1 and 0.4~$\mu$m thick YBCO films (Fig.~\ref{sem}(a) and (b), respectively). Notably, the holes are also visible on the surface of the thinnest (0.1~$\mu$m) sample presented (Fig.~\ref{sem}(a)), being of a much smaller diameter than for thicker films (the arrows in Fig.~\ref{sem}(d)).

Importantly, the surface structure presented for the multilayer is independent of the deposition temperature range presented in this work. This indicates that the improved smoothness results not from different (improved) growth conditions, but from the relaxed strains in the crystal lattice of YBCO by multilayering it with NdBCO. The crystal lattice strain relaxation is achieved through the creation of additional edge dislocations at the interfaces between YBCO and NdBCO layers due to the mismatch between their crystal lattices. Indeed, the ratios between crystal lattice parameters are quite similar for the YBCO/STO interface (for parameter $a$: 3.821~{\AA}$/3.905$~{\AA} $ = 0.9785$, for $b$: 3.885~{\AA}$/3.905$~{\AA} $ = 0.9949$) and for YBCO/NdBCO interfaces (for $a$: 3.821~{\AA}$/3.865$~{\AA} $ = 0.9886$, for $b$: 3.885~{\AA}$/3.9163$~{\AA} $ = 0.992$), which assume that all interfaces act in a similar way with a slightly larger number of edge dislocations \cite{svet} to be expected from STO/YBCO interface than from YBCO/NdBCO ones. In fact, not only NdBCO positively influences on YBCO structure formation, but also underlaying YBCO layers promote the simplified growth of the NdBCO layers. The YBCO layers act as seed layers for NdBCO growth in a similar way as described in Ref.~\cite{jia03}. As a result, the relaxed, ``seeded" crystal lattice experiences improved growth conditions for achieving high quality multilayers with large thicknesses, as evident from a wider temperature range for deposition.

\begin{figure}%[t]
%\begin{center}
\vspace{2.4cm}
\hspace{-0.5cm}
\includegraphics[width=90mm]{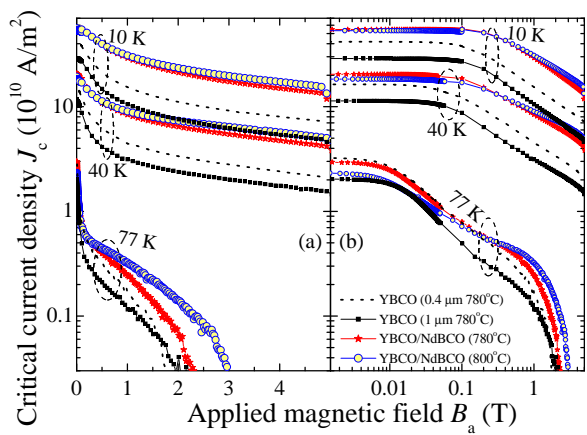}
\vspace{-2.7cm}
\caption{Critical current density as a function of applied magnetic field in (a) semi-logarithmic and (b) double-logarithmic scales. The 0.4 and 1 $\mu$m thick YBCO mono-layer films, as well as YBCO/NdBCO multilayers grown at different temperatures are shown for comparison.}
\label{jc}
%\end{center}
\end{figure}

The $J_c(B_a)$ dependences for some above-discussed samples are provided in Fig.~\ref{jc}. Two YBCO films: one of nearly optimal thickness (0.4~$\mu$m) and the other being 1~$\mu$m thick are presented for comparison. The 0.4~$\mu$m thick YBCO film shows $J_c(0) = 3.2 \times 10^{10}$~A/m$^2$ in zero field at $T = 77$~K. Strikingly, the multilayers outperform both YBCO films in the nearly entire field range. The only exception, where $J_c$ of the 0.4~$\mu$m thick film is slightly larger than that for the 1~$\mu$m multilayer film, is for $B_a < 0.3$~T at $T = 77$~K. The $J_c$ enhancement of 1~$\mu$m thick multilayer compared to the mono-layer film of the same thickness is by a factor of $\simeq 2$ at zero field and by a factor of $> 3$ at 1~T over the entire measured temperature range. As can be seen in Fig.~\ref{jc}, different deposition temperatures shown result in slightly different $J_c(B_a)$ curves for the multilayers. This indicates variation of defect structure formation with changing temperature and corresponding pinning properties.

We suggest that the effect of the drastic $J_c$ enhancement in the multilayers over entire field range is twofold. On one hand, the filling factor determined from SEM observation is about 10 to 18\% larger than for YBCO films, which produces the higher $J_c$ especially at low fields due to a larger effective cross-section for supercurrents. On the other hand, due to different ionic radii of Y and Nd, YBCO and NdBCO systems have a mismatch between crystal lattice parameters. This results in additional local stress near the YBCO/NdBCO interfaces, leading to formation of additional pinning sites in the form of out-of-plane edge dislocations. These extended defects are known to be one of the strongest pinning sites in YBCO films \cite{vpan}. Therefore, the formation of additional defects at the interfaces is likely to lead to the $J_c$ enhancement observed, in particular at large fields (as best seen in Fig.~\ref{jc}(a)). The strongest pinning of the edge dislocations is achieved at low temperatures where the radius of a vortex core $\xi(T) = \xi(0)(1-T/T_c)^{-1/2}$ is nearly equal to the radius of the dislocation core \cite{vpan}: $\xi(10 \,{\rm K}) \simeq 1.59$~nm. At $T = 77$~K, $\xi \simeq 4.01$~nm, which is more than a factor of two larger than the dislocation core, which implies weaker pinning. However, dislocations can still pin the vortices at higher temperatures since the superconducting order parameter is reduced around the dislocations due to elastic strains developed by the dislocations. Experimentally, a reduced $J_c$ enhancement for the multilayer is observed for 77~K compared to 10~K results (Fig.~\ref{thick}), indicating that the dislocations are likely to be responsible for the enhancement. Furthermore, the results obtained for the multilayers in which the thickness of the YBCO layers has been varied from 100 to 400~nm did not show any significant difference in terms of their $J_c(B_a)$ dependence, as should be expected for the out-of-plane dislocations, extending throughout the entire layer thickness.

The effect of larger surface supercurrent contribution to the total $J_c$ compared to the bulk supercurrents, producing larger overall $J_c$ in thinner films at low fields \cite{li04} is expected to be stronger at lower temperatures where the demagnetising effect is the most pronounced. This contradicts to our observations (Fig.~\ref{jc}), the 0.4~$\mu$m thick film shows the larger $J_c$ than the thicker multilayer at low fields only at 77~K. Therefore, the surface supercurrent contribution influence is ruled out in our work. The $J_c$ behavior at low fields is more likely to be governed by the structural peculiarities.

\section{Conclusion}

In summary, the YBCO/NdBCO multilayers have been shown to have {\em two dramatic improvements} compared to similar monolayer YBCO films. (i) The 1~$\mu$m thick YBCO/NdBCO multilayers with NdBCO layers comprising only 10\% of the entire film have been shown to outperform YBCO monolayer films with any thickness $\le 1 \, \mu$m in terms of $J_c(B_a)$ dependence. Possible reasons for the observed performance are considered to be a larger filling factor (less holes, smooth surface) and additional extended defects created in the multilayers at the YBCO/NdBCO interfaces as a result of crystal lattice mismatch between YBCO and NdBCO. (ii) The other important result obtained is the significantly improved smoothness of the film surface, which is also likely to be the result of the defect formation and corresponding strain release of the entire crystal structure in the multilayers.

\begin{acknowledgments}
This work is financially supported by the Australian Research Council.
\end{acknowledgments}

%\newpage

%\newpage

\end{document}